\documentclass[fleqn,12pt,twoside]{article}
\usepackage{espcrc1}

\usepackage{graphicx}
\usepackage{epsfig}
\usepackage[figuresright]{rotating}

\newcommand{\AmS}{{\protect\the\textfont2
  A\kern-.1667em\lower.5ex\hbox{M}\kern-.125emS}}

\title{High $p_T$ Measurements from PHENIX}

\author{S.~Mioduszewski\address{Physics Department
, 
	Brookhaven National Laboratory, Upton, NY 11973, USA} for the PHENIX Collaboration\thanks{for the full PHENIX Collaboration author list and acknowledgements, see Appendix ``Collaborations'' of this volume.}}
       
\begin{document}

\maketitle

\begin{abstract}
We present recent high transverse momentum measurements 
by the PHENIX experiment for Au+Au and p+p collisions 
at $\sqrt{s_{NN}} = 200$~GeV at the Relativistic Heavy Ion Collider (RHIC).  
We show particle spectra for neutral pions and charged hadrons, 
define and show the nuclear modification factor, 
and discuss particle composition.  By means of the nuclear modification factor,
we observe a suppression factor of 5-6 for neutral pions and 3-4 for charged
hadrons in central collisions at high $p_T$.
We find that the ratio of $\pi^0$ to $(h^++h^-)/2$ 
remains nearly constant at $\sim 0.5$ for $p_T = 2-9$~GeV/c.
Finally we present strong evidence for the observation 
of jets in Au+Au collisions.
\end{abstract}

\section{Introduction}

In Run~II at RHIC ($\sqrt{s_{NN}} = 200$~GeV), PHENIX has been able to 
measure particle yields up to large transverse momenta, $p_T \sim 10$~GeV/c,
where the cross section is expected to be dominated by hard processes.  
This high statistics data set allows us to study the systematics 
of high $p_T$ phenomena in detail and at much higher $p_T$ than possible in Run~I ($\sqrt{s_{NN}} = 130$~GeV) data.

In p+p collisions, it is known that hard scattering and the fragmentation
of the scattered partons dominates the production of
hadrons above $p_T \sim 2$~GeV/c~\cite{Owens}.  In Au+Au collisions, these hard scatterings are of particular interest because they occur early in the collision, leaving the hard-scattered partons sensitive to the properties of the collision medium. 
Further interactions of these partons in the dense medium 
might cause the high transverse momentum tail of the hadron spectrum, 
where the hadrons are likely to be the leading particles of jets, to be 
suppressed~\cite{quench}. Due to the high multiplicity
environment, jets cannot be directly observed in a heavy ion collision.
We can however determine the contribution of jet fragmentation
to hadron yields at high $p_T$ via correlation measurements and
use this knowledge to interpret the measured hadron $p_T$ spectra.  

First, we give a brief overview of the PHENIX detector.  
In the following sections,
we proceed assuming that the hadrons at high $p_T$ emanate from jets, with
the objective of measuring the effect of the dense medium on the
hard-scattered partons through modification of the particle spectra 
at high $p_T$.  We define the nuclear modification factor and show such
a quantity for both neutral pions and charged hadrons.  With the measurement 
of spectra for both charged hadrons and neutral pions, 
we also investigate the particle composition at high $p_T$. 
Finally, we use two-particle correlations to test the assumption
that the high $p_T$ hadrons are due to the fragmentation of hard-scattered partons.

\section{Overview of the PHENIX Detector}

PHENIX is a versatile detector, with tracking chambers, ring-imaging 
Cerenkov (RICH) detectors, a time-of-flight (TOF) wall, and electromagnetic 
calorimeters in the central arms, which cover $\Delta \eta = \pm 0.35$ and 
$180^o$ in azimuth.  Along the beamline there are beam-beam counters and 
zero-degree calorimeters, which are used for event triggering and centrality 
classification.  Neutral pions are measured via their 
$\gamma\gamma-$decay using the electromagnetic calorimeters (EMCal).  
There are 6 sectors of Lead-Scintillator (PbSc) and 2 sectors 
of Lead-Glass (PbGl) calorimeter, providing 
two independent measurements of the $\pi^0$~$p_T$ spectrum.  
The charged hadron spectrum is measured using the tracking 
detectors: drift chambers and pad chambers.  
At $p_T$ above $\sim 5$~GeV/c the RICH detector, 
together with an energy cut in the PbSc, is used to identify charged pions.  
With various detection methods, PHENIX can make multiple overlapping 
measurements at high $p_T$, providing valuable consistency checks.
Details of the detector are described elsewhere~\cite{detector}.  

\section{High $p_T$ Particle Spectra}

\begin{figure}[!htb]
\begin{center}
\vspace{-0.6cm}
\epsfig{figure=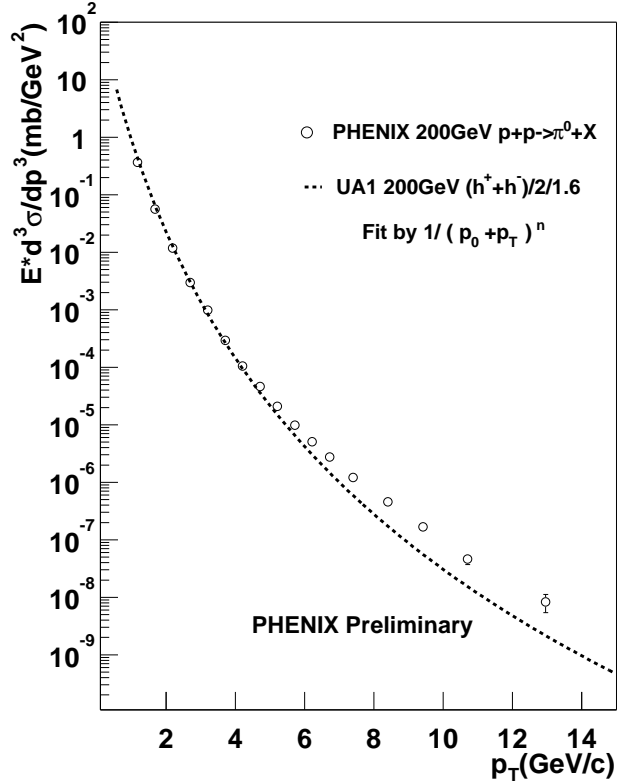,height=4.5in}
\vspace{-1.2cm}
\caption{Neutral pion production cross section vs. $p_T$ in p+p collisions at
$\sqrt{s_{NN}} = 200$~GeV.\label{pi0_pp}}
\vspace{-0.3cm}
\end{center}
\end{figure}

In Run~II, PHENIX recorded more than 30~million minimum bias Au+Au events.  
This event sample enables us to reach 10~GeV/c in the transverse momentum 
spectra of both identified neutral pions and charged hadrons.  In addition,
we recorded 140~million p+p events in which we measure neutral pions up
to 13~GeV/c.  Shown in Fig.~\ref{pi0_pp} is the neutral pion production 
cross section as a function of $p_T$, 
measured with the PbSc, for p+p collisions~\cite{torii}.  
The measured p+p spectrum is compared to a fit to 
UA1 data~\cite{UA1} over the range $p_T < 6$~GeV/c, 
extrapolated to higher $p_T$.  Although our measurement agrees with the UA1 
data, it disagrees with this extrapolation 
at very high $p_T$.  This new measurement provides an important 
reference spectrum for comparison with the Au+Au data.
 
\begin{figure}[!htb]
\begin{minipage}[t]{7.5cm}
\vspace{-0.8cm}
\epsfig{figure=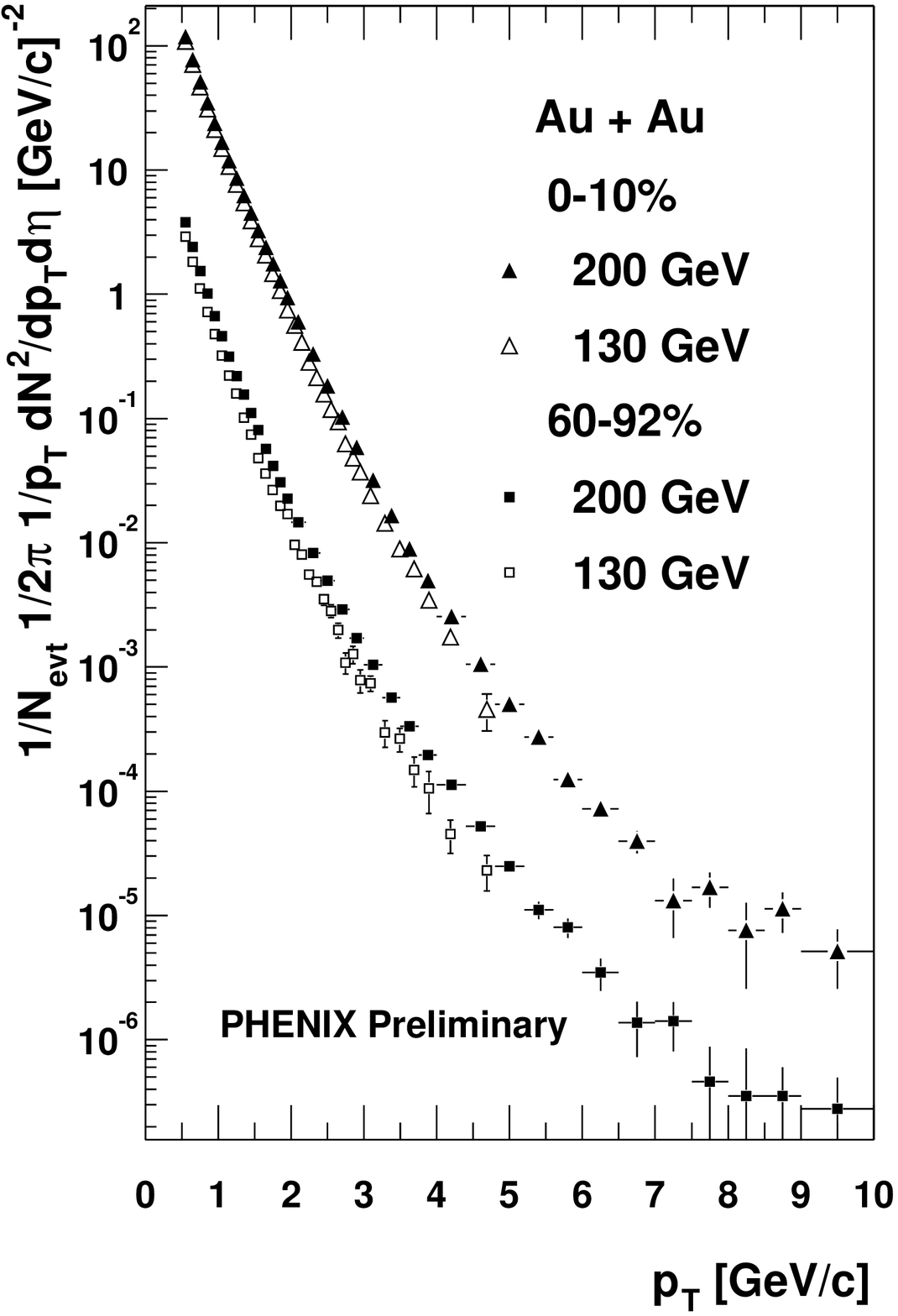,height=4.8in}
\vspace{-1.4cm}
\caption{Invariant yields of $(h^++h^-)/2$ 
as a function of $p_T$ in Au+Au collisions at
$\sqrt{s_{NN}} = 200$~GeV~\cite{jjia} and 130~GeV for centrality 
selections of 0-10\% and 60-92\%.}
\vspace{-0.3cm}
\label{chg_auau}
\end{minipage}\hfill
\begin{minipage}[t]{7.5cm}
\vspace{-0.2cm}
\hspace{-0.8cm}
\epsfig{figure=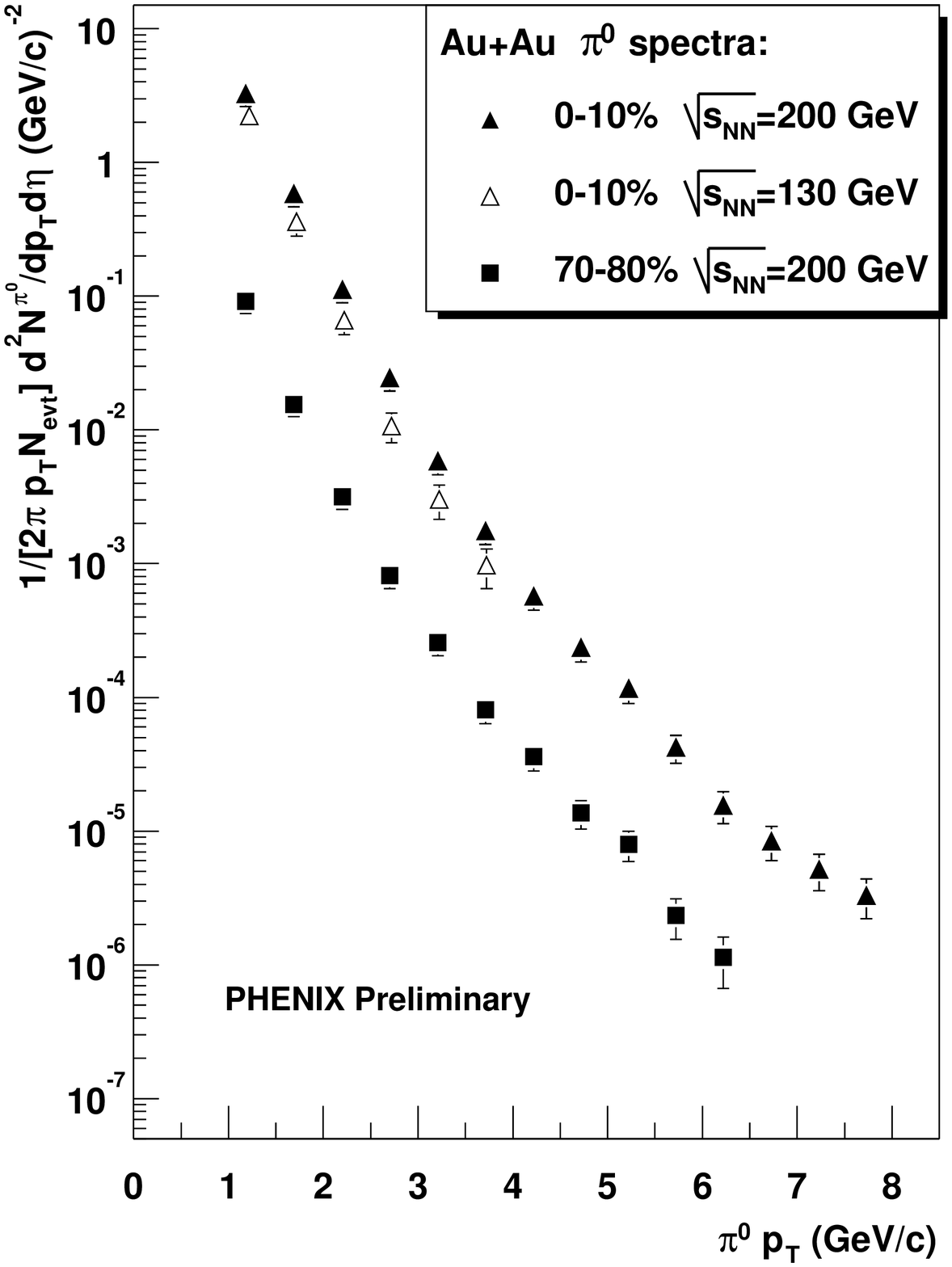,height=4.4in}
\vspace{-0.94cm}
\caption{Invariant yields of $\pi^0$ 
as a function of $p_T$ in Au+Au collisions at
$\sqrt{s_{NN}} = 200$~GeV and 130~GeV for centrality 0-10\% and 
$\sqrt{s_{NN}} = 200$~GeV for centrality 70-80\%.}
\vspace{-0.3cm}
\label{pi0_auau}
\end{minipage}
\end{figure}

For Au+Au collisions, the events are binned into centrality
selections, and the spectra are shown for the most central bin and a
peripheral bin in Figs.~\ref{chg_auau} and~\ref{pi0_auau}. 
Centrality is expressed as a percentage of the total inelastic cross section.
In the Run~II data, the charged hadron yields are measured up to
$p_T = 10$~GeV/c for central events and $p_T = 5$~GeV/c for
peripheral events, while the neutral pion yields 
are measured up to 8 and 6.5~GeV/c 
for central and peripheral events respectively.

\section{$R_{AA}$ for Neutral Pions}

\begin{figure}[!htb]
\begin{minipage}[t]{7.5cm}
\vspace{-0.6cm}
\hspace{-0.2cm}
\epsfig{figure=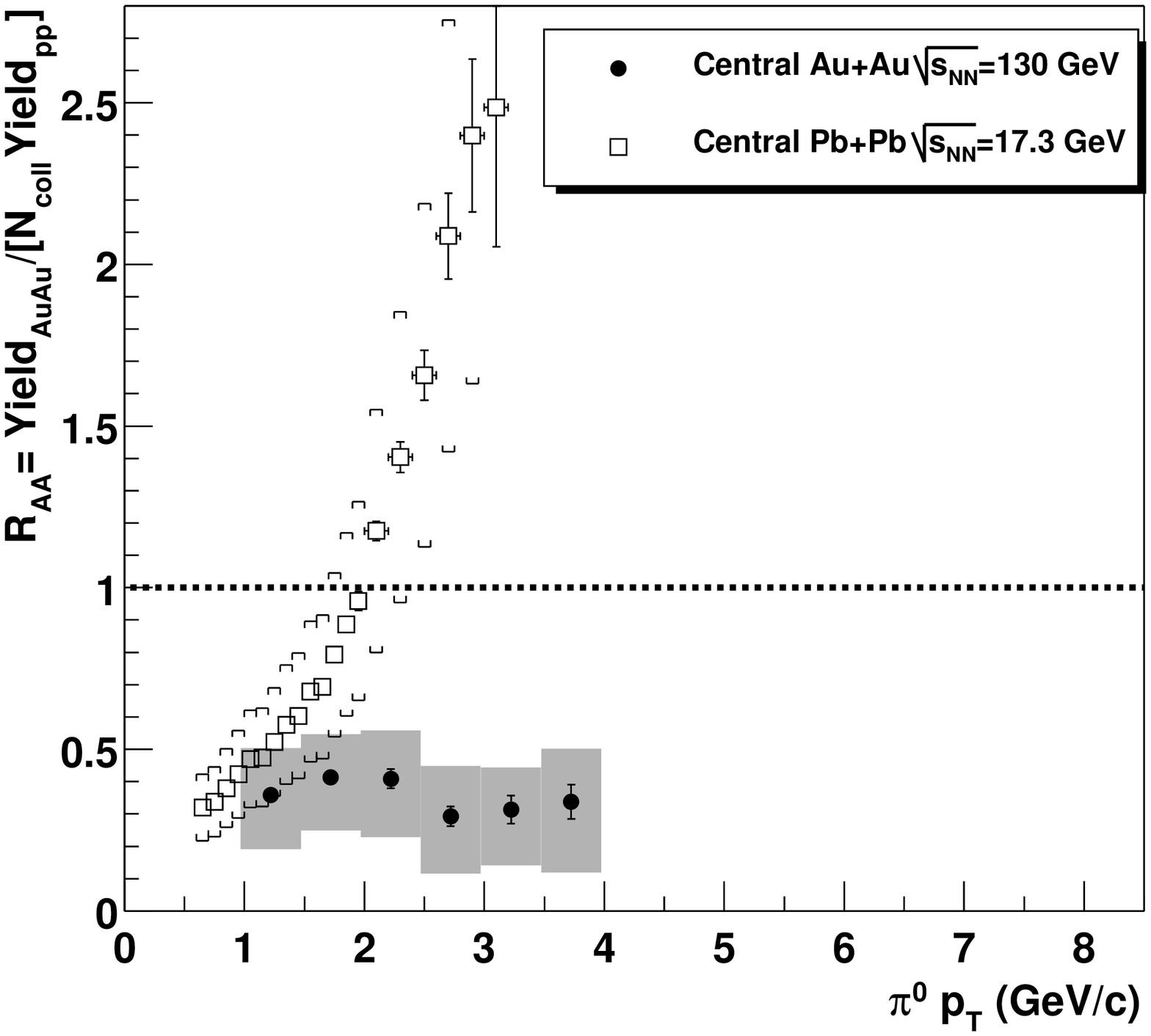,height=3.1in}
\vspace{-1.2cm}
\caption{$R_{AA}$ ($\pi^0$) 
for central Pb+Pb collisions at $\sqrt{s_{NN}}=17$~GeV and 
central Au+Au collisions at $\sqrt{s_{NN}}=130$~GeV.  The error bars are the
statistical $\oplus$~$p_T$-dependent systematic errors.  
The brackets/boxes are the errors on the normalization of this ratio.}
\vspace{-0.3cm}
\label{raa_ecomp}
\end{minipage}\hfill
\begin{minipage}[t]{7.5cm}
\vspace{-0.6cm}
\hspace{-0.6cm}
\epsfig{figure=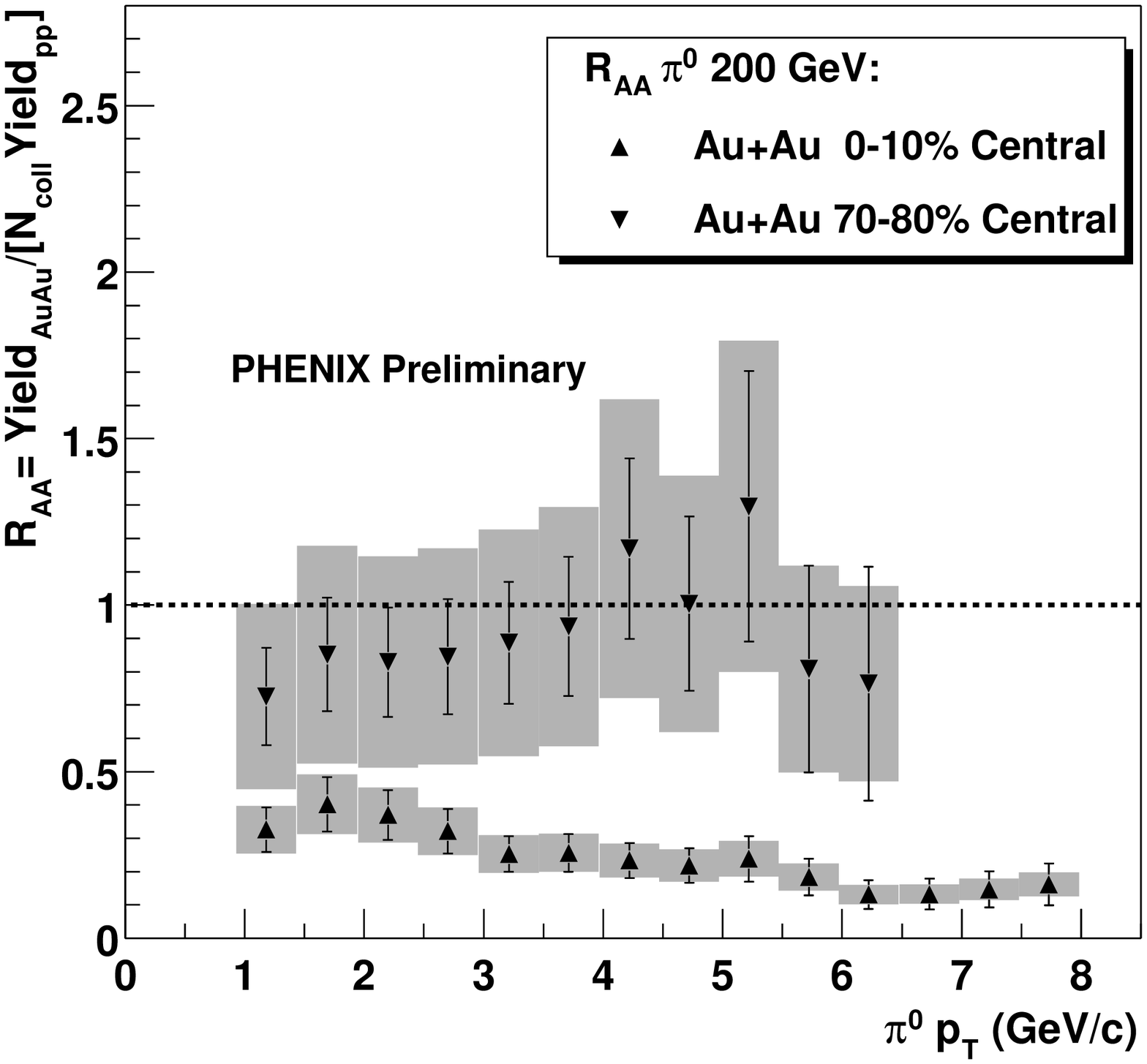,height=3.1in}
\vspace{-1.2cm}
\caption{$R_{AA}$ ($\pi^0$) for central and peripheral Au+Au collisions at 
$\sqrt{s_{NN}}=200$~GeV.  The error bars are the
statistical $\oplus$~$p_T$-dependent systematic errors.  
The shaded boxes are the errors on the normalization of the ratio.}
\vspace{-0.3cm}
\label{raa_periph}
\end{minipage}
\end{figure}

The nuclear modification factor quantifies the effect of A+A compared to p+p
collisions on particle yields for point-like processes.  
It is defined as the ratio of the particle yield 
in a A+A collision to the yield in a p+p collision scaled by the mean number
of binary (nucleon+nucleon) collisions $N_{coll}$ in the A+A event sample.
Because hard processes are generally believed to scale with $N_{coll}$ 
(``binary-scaling''), 
this ratio is expected to be one at high $p_T$ in the absence 
of any nuclear effects.
\begin{equation}
R_{AA}(p_T) 
= \frac{(\rm Yield \; per \; A+A \; collision)}
{\langle N_{coll} \rangle ({\rm Yield \; per \; p+p \; collision})}
= \frac{ d^{2}N^{A+A}/dp_T d\eta }
{\langle N_{coll} \rangle  
( d^{2}\sigma^{p+p}/dp_T d\eta ) / \sigma^{p+p}_{inelastic} }.
\label{eq:RAA_defined}
\end{equation}
Shown in Fig.~\ref{raa_ecomp} is $R_{AA}$ as a function of $p_T$ for
central Pb+Pb collisions at $\sqrt{s_{NN}}=17$~GeV from WA98~\cite{WA98} and
central Au+Au collisions at $\sqrt{s_{NN}}=130$~GeV 
from PHENIX Run~I~\cite{PHENIX_supp}.
Already in this comparison, the suppression relative to binary-scaling 
at 130~GeV is striking.  
It is quite different from the enhancement observed at 17~GeV.  
This enhancement is known as the ``Cronin effect''~\cite{Cronin} and has been 
attributed to $p_T$ broadening due to initial state scatterings~\cite{ptbroadening,pre_constantE}.
Figure~\ref{raa_periph} shows $R_{AA}$ for central and peripheral Au+Au 
collisions at $\sqrt{s_{NN}}=200$~GeV.
The central data again show a suppression as seen at 130~GeV.  With 
a measurement that extends to much larger $p_T$, the suppression is shown to 
persist up to $p_T \sim 8$~GeV/c
and is as much as a factor of 5-6 at the largest $p_T$.  
The measurement of the p+p reference spectrum with the same detector
greatly reduces the systematic error in $R_{AA}$ for central events.  
The $R_{AA}$ for peripheral events is consistent with binary-scaling 
(p+p yields scaled
by the number of binary collisions in the peripheral event sample).  
The error in this ratio is dominated by the uncertainty in 
$\langle N_{coll} \rangle$ 
for peripheral collisions.  Within the rather large errors, 
the spectrum for peripheral collisions does not show effects of the nuclear
medium.  If there is a Cronin effect at RHIC, it is not strongly evident in
peripheral collisions.

\begin{figure}[!htb]
\begin{center}
\vspace{-0.8cm}
\epsfig{figure=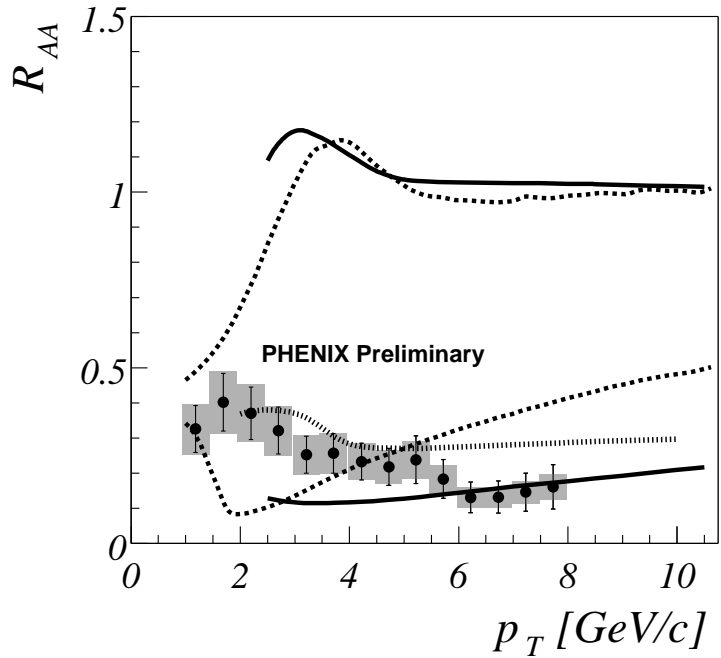,height=4.25in}
\vspace{-2.0cm}
\caption{$R_{AA}$ as a function of $p_T$ compared to theoretical predictions
for central Au+Au collisions at $\sqrt{s_{NN}} = 200$~GeV.  The dotted 
lines are
predictions with and without energy loss (constant)~\cite{pre_constantE}.  
The solid lines are again predictions with and without energy loss (energy-dependent)~\cite{pre_nonconstantE1}, and the hashed line is also a prediction with energy-dependent energy loss~\cite{pre_nonconstantE2}. \label{raa_theory}}
\vspace{-0.6cm}
\end{center}
\end{figure}

There are a number of model predictions, shown in Fig.~\ref{raa_theory}, 
that were made prior to the release
of the Run~II data.  In the case of one model which assumes energy loss 
that is constant with respect to the parton's 
energy~\cite{pre_constantE}, $R_{AA}$ increases by a factor
of $\sim 2$ from $p_T \sim 3.5$ to $p_T \sim 8$~GeV/c.  Although
the magnitude of the suppression qualitatively describes the data at 
$p_T \sim 4$~GeV/c, the increasing $R_{AA}$ with $p_T$ is contrary to what
is observed in central Au+Au collisions.
Two other models which use a common formalism for including energy-dependent 
energy loss differ in the point where the
calculation is begun (thus resulting in somewhat different predictions).
One such model predicts an $R_{AA}$ that increases only slightly 
with increasing $p_T$~\cite{pre_nonconstantE1}, while another model 
predicts a nearly constant $R_{AA}$ from $p_T \sim 4$ to at least 
$p_T \sim 10$~GeV/c~\cite{pre_nonconstantE2}.
The energy-dependent energy loss models give reasonable agreement with 
the data.  This is also seen in another calculation 
made subsequent to the release of the Run~II data~\cite{postdiction}.

\section{Comparison of Central Collisions with Respect to Peripheral Collisions}

\begin{figure}[!hbt]
\begin{center}
\vspace{-0.8cm}
\epsfig{figure=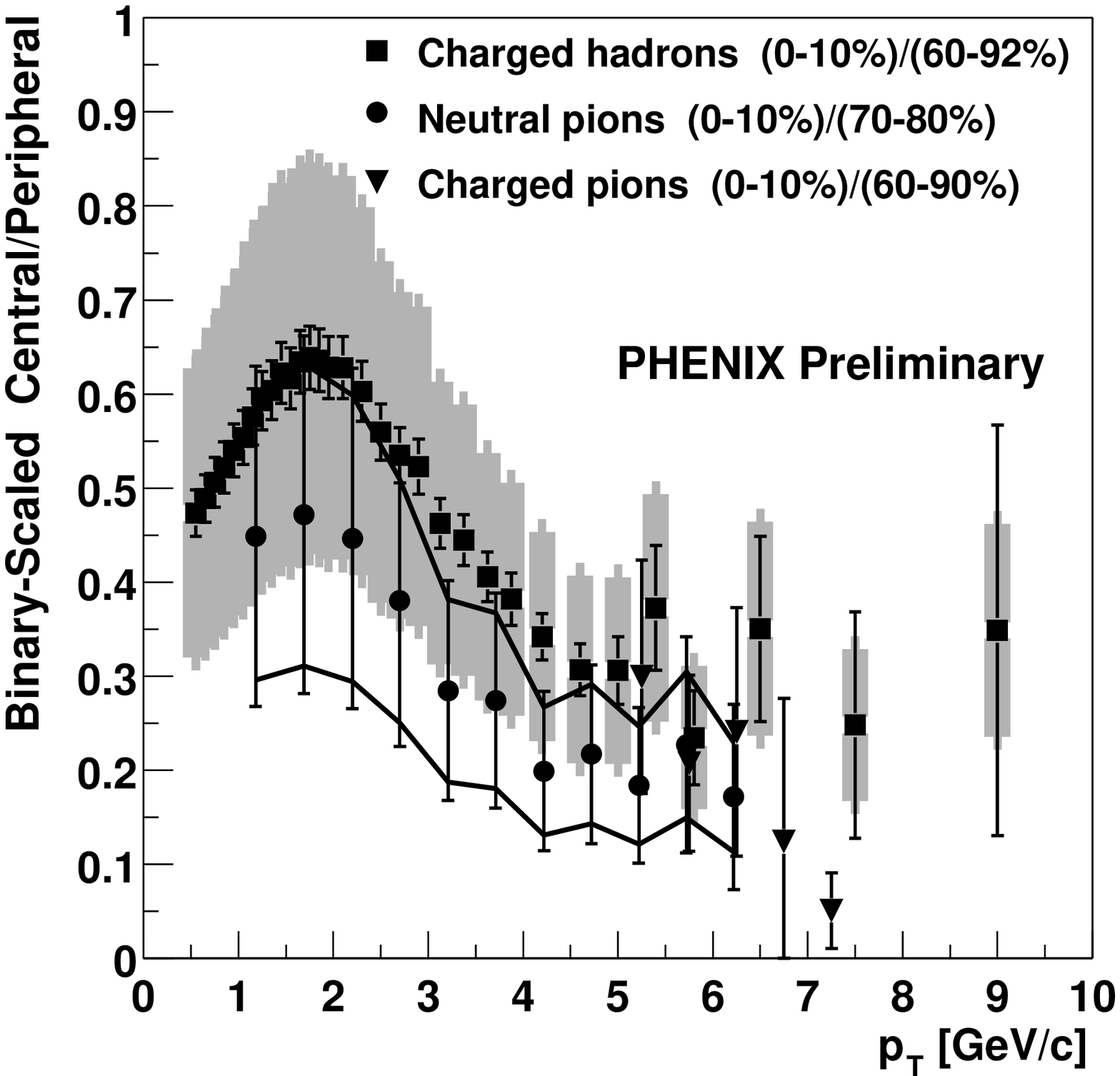,height=4.0in}
\vspace{-0.6cm}
\caption{Binary-scaled central-to-peripheral ratio vs. $p_T$ 
for charged hadrons, neutral pions, and charged pions measured in Au+Au 
collisions.  The error bars are
statistical $\oplus$~$p_T$-dependent systematic errors.  The error in the 
overall 
normalization is shown as outlines for neutral pions and a shaded area for 
charged hadrons.  For charged pions, the error is dominated by statistics.\label{pi0_charged}}
\vspace{-0.3cm}
\end{center}
\end{figure}

An alternate measure of the nuclear modification is the
central-to-peripheral ratio scaled by the number of binary collisions, 
\begin{equation}
\rm Binary\rm{-}Scaled \; Central/Peripheral = \frac{(\rm Yield \; per \; central \; collision)/\langle N_{coll}^{central} \rangle }
{({\rm Yield \; per \; peripheral \; collision})/\langle N_{coll}^{peripheral} \rangle}. 
\end{equation} 
Since the $\pi^0$ spectrum in peripheral Au+Au
collisions is consistent with the binary-scaled p+p $\pi^0$ spectrum 
(Fig.~\ref{raa_periph}), this ratio is similar to $R_{AA}$.  
An advantage of this
ratio is that many of the systematic uncertainties in the measurement
cancel, particularly for charged hadrons.  A disadvantage is that it is 
sensitive to the centrality dependence of the Cronin effect, 
which is not known.
Figure~\ref{pi0_charged} shows this ratio as a function of $p_T$ for 
neutral pions, charged hadrons, and charged pions.
The suppression is observed in all three measurements.  
In the charged hadrons, the
suppression factor reaches 3-4 at high $p_T$.  The difference in this
ratio between the identified pions and charged hadrons seems to be due to
the particle composition.  In particular, we find that
$p/\pi^0 \sim 1$ in central collisions for $p_T = 2-4$~GeV/c while
$p/\pi^0 \sim 0.4$ in peripheral collisions~\cite{takao}. 
The error in the normalization, denoted by a shaded area for the charged 
hadrons and outlines for the neutral pions, is dominated by the uncertainty
in $\langle N_{coll} \rangle$ for peripheral collisions.

\section{Centrality Dependence of $R_{AA}$}

\begin{figure}[!htb]
\begin{minipage}[t]{7.5cm}
\vspace{-0.6cm}
\hspace{-0.32cm}
\epsfig{figure=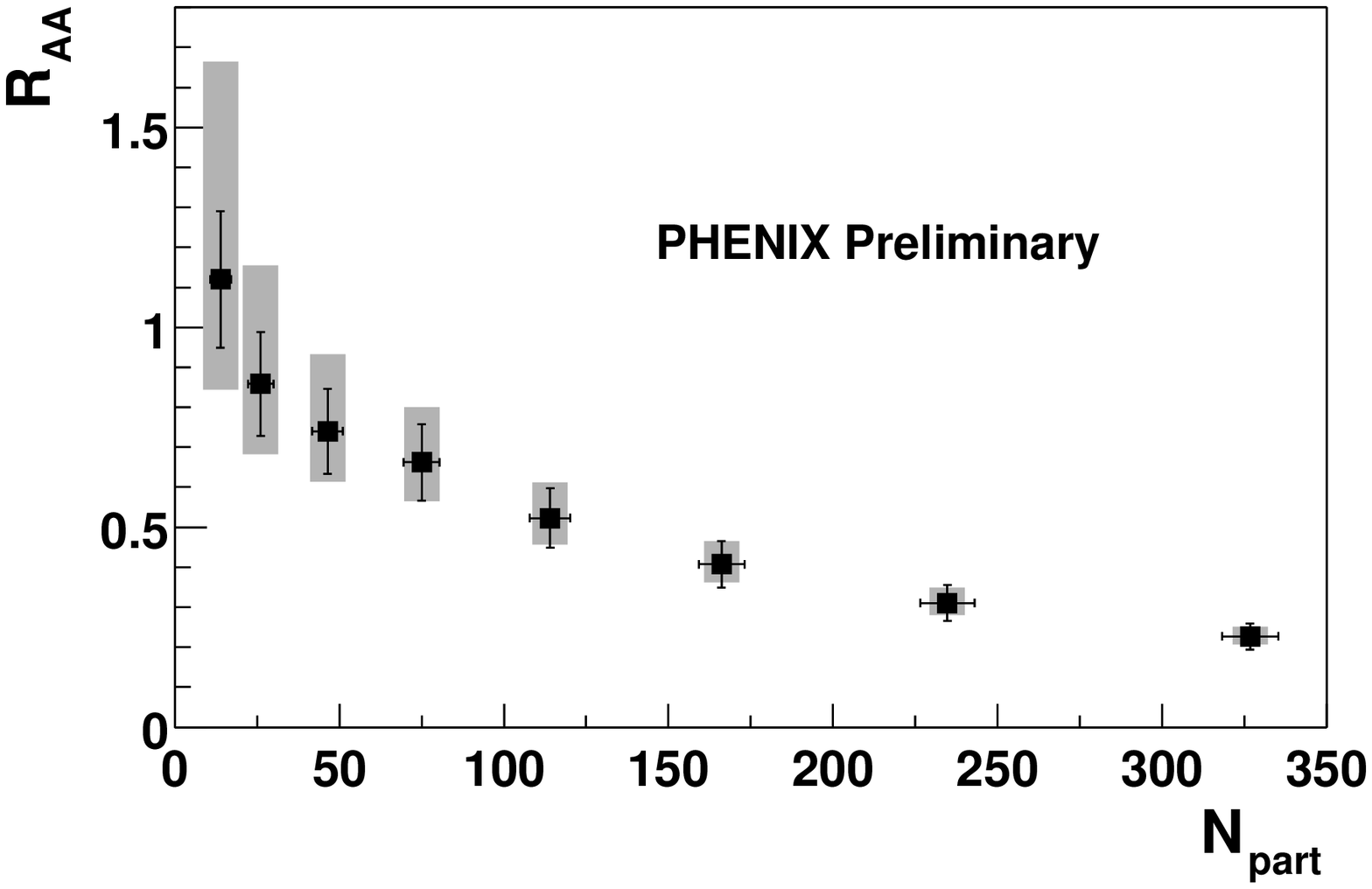,height=2.3in}
\vspace{-1.2cm}
\caption{$R_{AA}$ vs. $N_{part}$ (or centrality of the collision) for neutral 
pions with $4 < p_T < 6$~GeV/c.  The shaded boxes denote the error on 
$\langle N_{coll}\rangle$.  There is an additional error of 20\% in the overall normalization that is not shown.}
\vspace{-0.3cm}
\label{raa_cent}
\end{minipage}\hfill
\begin{minipage}[t]{7.5cm}
\vspace{-0.6cm}
\hspace{-0.7cm}
\epsfig{figure=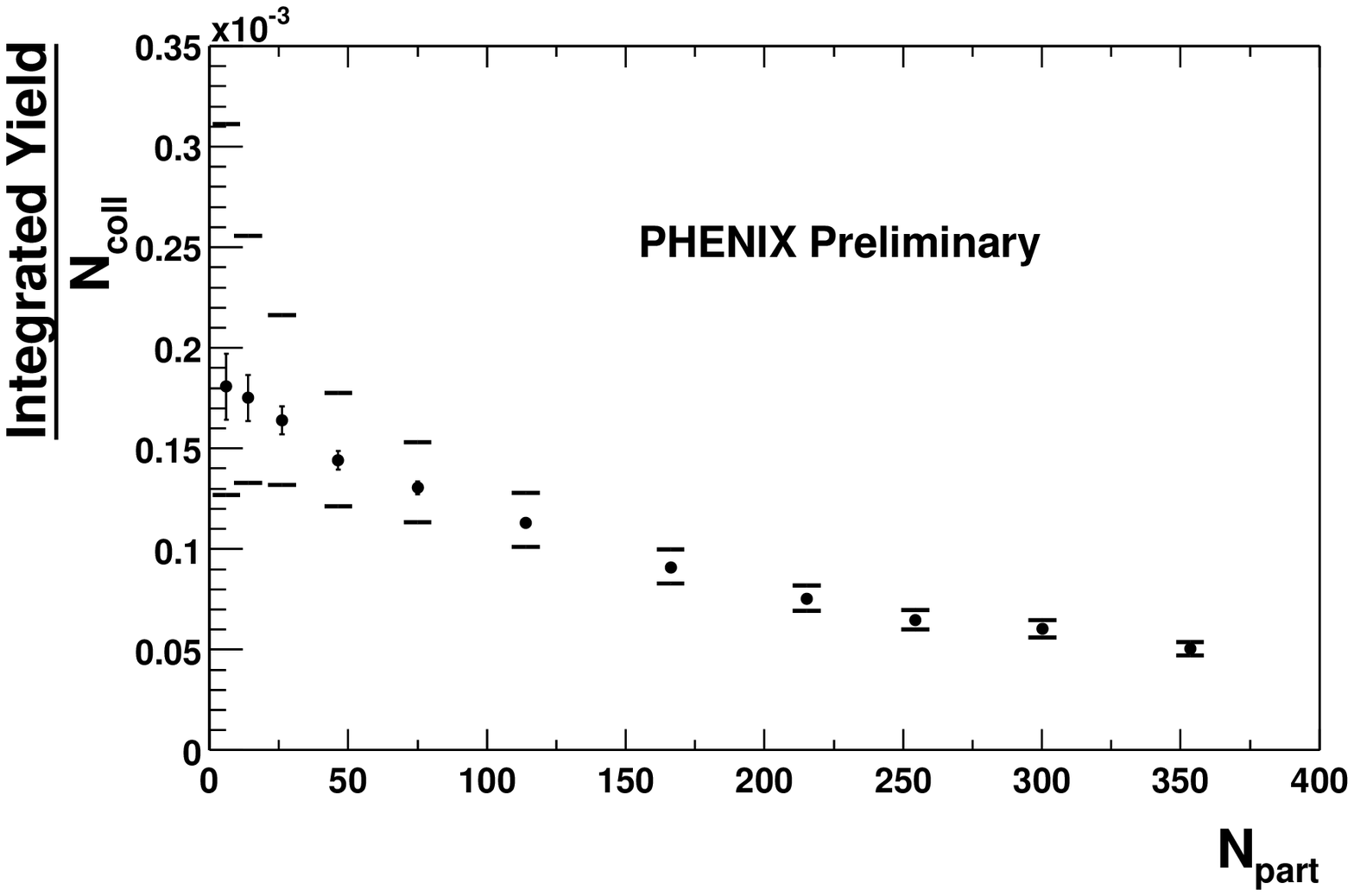,height=2.3in}
\vspace{-1.2cm}
\caption{Integrated yield scaled by $N_{coll}$ vs. $N_{part}$ for charged 
hadrons with $p_T > 4$~GeV/c.  The brackets denote the error on 
$\langle N_{coll}\rangle$.}
\vspace{-0.3cm}
\label{chg_cent}
\end{minipage}
\end{figure}

As shown in Fig.~\ref{raa_periph}, the 10\% most central events have a 
suppression factor reaching 5-6, while the 70-80\% centrality selection is
consistent with binary-scaled p+p collisions.  We now address the evolution of
the nuclear modification factor from the most peripheral to the 
most central collisions, in terms of the
mean number of participating nucleons $N_{part}$ in a centrality selection.  
For the 10\% most central collisions, $\langle N_{part}\rangle = 327$, 
and for the
70-80\% central collisions, $\langle N_{part}\rangle = 14$.
Figure~\ref{raa_cent} shows $R_{AA}$ vs. $N_{part}$ for
neutral pions with $4 < p_T < 6$~GeV/c.  
The suppression gradually increases from
peripheral to central events.  The same trend is seen in the charged hadron
yields.  Figure~\ref{chg_cent} shows the integrated yield scaled by $N_{coll}$
vs. $N_{part}$ for charged hadrons.  This quantity is similar to $R_{AA}$,
differing only in normalization by the p+p reference yields.
Again the binary-scaled yield decreases, or the suppression increases, 
gradually with increasing $N_{part}$.

\section{Particle Composition at High $p_T$ \label{sec:part_comp}}

\begin{figure}[!htb]
\begin{center}
\vspace{-0.6cm}
\epsfig{figure=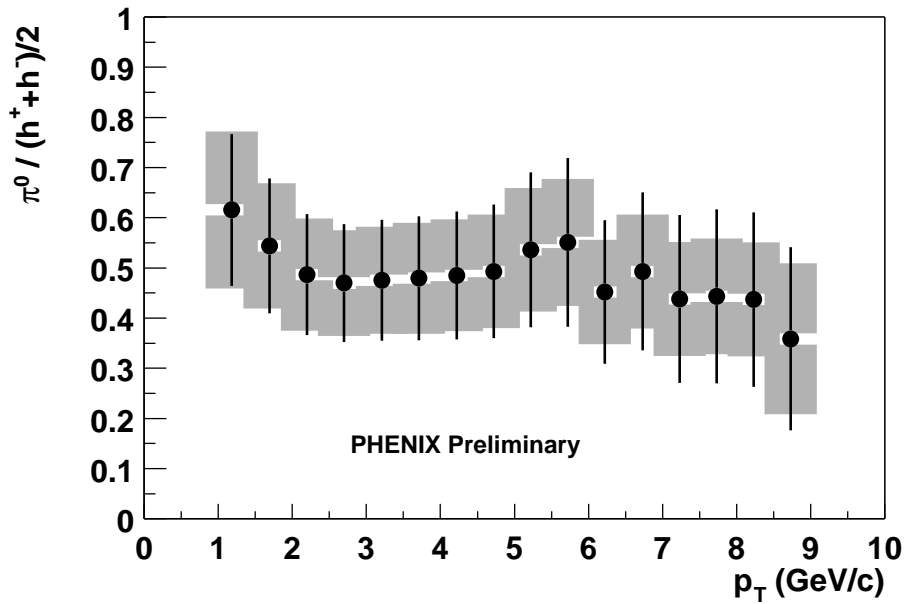,height=3.5in}
\vspace{-1.2cm}
\caption{Ratio of $\pi^0$ to $(h^++h^-)/2$ as a function of $p_T$ for minimum
bias Au+Au collisions.\label{pi_to_h}}
\vspace{-0.3cm}
\end{center}
\end{figure}

PHENIX can measure the $p/\pi$ ratio up to almost 4~GeV/c in 
$p_T$~\cite{takao}, 
at which point protons can no longer be distinguished from
other particles via their time of flight.  Since neutral pions are 
identified to much higher $p_T$, we can look at the ratio of $\pi^{0}$ to
nonidentified charged hadrons $(h^{+}+h^{-})/2$ for $p_T > 4$~GeV/c.  
This ratio is shown in Fig.~\ref{pi_to_h} for minimum bias events.
The surprising feature that this ratio does not increase 
for transverse momenta greater than 4~GeV/c, but remains nearly constant at
a value around 0.5, indicates that approximately half of the charged hadrons at
high $p_T$ are protons and/or kaons, assuming $\pi^0 = (\pi^++\pi^-)/2$.  
This result is rather different from hard processes as
measured in e$^+$e$^-$ collisions~\cite{Delphi}.  The interpretation for this
new result is not yet clear.  There may be 
some other production mechanism for protons and/or kaons in Au+Au collisions 
at these large transverse momenta.  Alternatively, the large proton 
and/or kaon content at high $p_T$
could be due to a difference in the suppression of pions relative to 
protons/kaons.

\section{Evidence for Jets}

\begin{figure}[!htb]
\begin{center}
\vspace{-0.6cm}
\hspace{-0.7cm}
\epsfig{figure=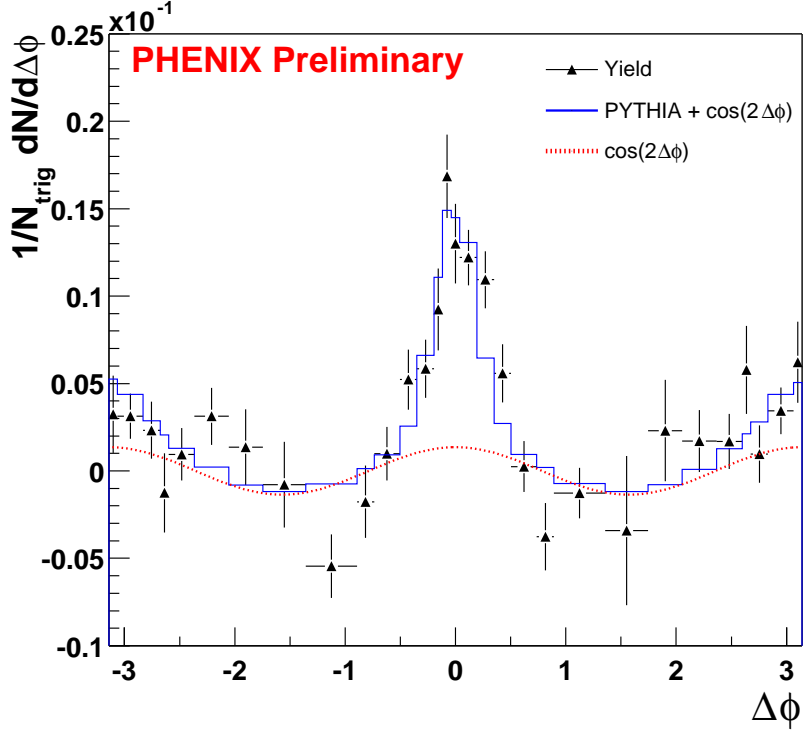,height=4in}
\vspace{-1.2cm}
\caption{Acceptance-corrected yield (in 20-40\% centrality selection) fit to PYTHIA plus an elliptic flow term.}
\vspace{-0.3cm}
\label{auau_dphi}
\end{center}
\end{figure}

Since jets cannot be directly observed in a Au+Au collisions, a correlation 
analysis is required to detect their presence.
We take a high $p_T$ neutral particle, a cluster in the EMCal with 
energy greater than 2.5~GeV, as the trigger and correlate it with all 
charged particles in the event within a given range in $p_T$.   
Comparing the correlations that we measure in p+p collisions at 
$\sqrt{s_{NN}} = 200$~GeV to those 
from the PYTHIA event generator~\cite{PYTHIA}, we establish 
that PYTHIA reproduces the behavior of jets known to be 
present in p+p collisions.  PYTHIA is then compared to Au+Au collisions
at $\sqrt{s_{NN}} = 200$~GeV, which demonstrates that the correlations we
observe in Au+Au collisions behave as expected for jets, both in the azimuthal 
and polar angles.  
Figure~\ref{auau_dphi} shows the background-subtracted correlation 
in the azimuthal angle $\Delta \phi$ of the trigger particle 
with charged particles having $p_T$ between 2 and 4~GeV/c.  
Also shown is a fit to the data of the correlation produced by PYTHIA with an added $cos(2\phi)$ term.  The $cos(2\phi)$ term 
accounts for the contribution of flow to the correlation, while the 
coefficient of the PYTHIA term represents the contribution due to jets.
Further details of this analysis are given elsewhere~\cite{mickey}.
From the figure, one can see that much of the 
correlation can be attributed to jets (PYTHIA), particularly the near-side 
correlation around $\Delta \phi = 0$.  The trigger particle is 
typically a photon from a $\pi^0$ decay.  This is strong evidence 
that when we trigger on a high $p_T$~$\pi^0$, we are indeed looking at a 
leading particle from a jet.

\section{Conclusions}

We have presented $p_T$ spectra for charged hadrons in Au+Au 
collisions and neutral pions in Au+Au and p+p collisions and have 
shown that the high $p_T$ suppression observed in Run~I~RHIC data at 
$\sqrt{s_{NN}} = 130$~GeV~\cite{PHENIX_supp,STAR_supp} 
persists up to $p_T \sim 8$~GeV/c at $\sqrt{s_{NN}} = 200$~GeV.
The neutral pions are suppressed by a factor of 5-6 and the charged hadrons by 
3-4 for $p_T > 4$~GeV/c in the most central collisions.  
The suppression increases gradually with increasing centrality, or $N_{part}$.
We have also shown that when 
triggering on a high $p_T$ neutral particle, which is
predominantly a photon from a $\pi^0$ decay, the correlation between the 
neutral and charged particles in the event is jet-like.  This is strong 
evidence that the neutral pions 
that we measure at high $p_T$ indeed have significant 
contributions from jet fragmentation.  Finally, we have shown that 
approximately half of the charged hadrons at high $p_T$ are not pions, but protons and/or kaons, up to $p_T \sim 9$~GeV/c.

\def\IJMPA{{Int. J. Mod. Phys.}~{\bf A}}
\def\JPG{{J. Phys}~{\bf G}}
\def\NCA{Nuovo Cimento}
\def\NIM{Nucl. Instrum. Methods}
\def\NIMA{{Nucl. Instrum. Methods}~{\bf A}}
\def\NPA{{Nucl. Phys.}~{\bf A}}
\def\NPB{{Nucl. Phys.}~{\bf B}}
\def\PLB{{Phys. Lett.}~{\bf B}}
\def\PLC{Phys. Repts.\ }
\def\PRL{Phys. Rev. Lett.\ }
\def\PRD{{Phys. Rev.}~{\bf D}}
\def\PRC{{Phys. Rev.}~{\bf C}}
\def\ZPC{{Z. Phys.}~{\bf C}}
\def\EPJC{{Eur.Phys.J.}~{\bf C}}


\begin{thebibliography}{9}
\bibitem{Owens} J.F.~Owens {\it et al.}, \PRD{\bf 18}, 1501 (1978).
\bibitem{quench} M.~Gyulassy and M.~Pl\"umer, \PLB{\bf 243}, 432
(1990); R.~Baier {\it et al.}, \PLB{\bf 345}, 277 (1995);  X.N.~Wang and M.~Gyulassy, \PRL{\bf 68}, 1480 (1992); X.N.~Wang, \PRC{58}, 2321 (1998).
\bibitem{detector} PHENIX Collaboration, D.~Morrison, {\it et al.},
\NPA{\bf 638}, 565c (1998);
PHENIX Collaboration, W.~Zajc, {\it et al.},
Quark Matter 2001.
\bibitem{torii} H.~Torii for the PHENIX Collaboration, these proceedings.
\bibitem{UA1} C.~Albajar {\it et al.}, \NPB{\bf 335}, 261 (1990).
\bibitem{jjia} J.~Jia for the PHENIX Collaboration, these proceedings.
\bibitem{enterria} D.~D'Enterria for the PHENIX Collaboration, these proceedings.
\bibitem{WA98} M.M.~Aggarwal {\it et al.} [WA98 Collaboration], Eur. Phys. J. C23 (2002) 225-236.
\bibitem{PHENIX_supp} K.~Adcox {\it et al.} [PHENIX Collaboration],
Phys.\ Rev.\ Lett.\  {\bf 88}, 022301 (2002).
\bibitem{Cronin} D.~Antreasyan {\it et al.}, \PRD{\bf 19}, 764 (1979).
\bibitem{ptbroadening} M.~Lev and B.~Petersson, Z. Phys. C{\bf21}, 155 (1983); T.~Ochiai {\it et al.}, Prog. Theor. Phys. 75, {\bf 288} (1986).
\bibitem{pre_constantE} X.N.~Wang, Phys. Rev. C61, 064910 (2000).
\bibitem{pre_nonconstantE1} P.~Levai {\it et al.}, Nuclear Physics A698 (2002) 631.
\bibitem{pre_nonconstantE2} I.~Vitev and M.~Gyulassy, these proceedings, hep-ph/0208108; M.~Gyulassy, P.~Levai and I.~Vitev, Nucl. Phys. B 594, p. 371 (2001). 
\bibitem{postdiction} S.~Jeon, J.~Jalilian-Marian and I.~Sarcevic, nucl-th/0208012.
\bibitem{takao} T.~Sakaguchi for the PHENIX Collaboration, these proceedings.
\bibitem{Delphi} P.~Abreu {\it et al.} [DELPHI Collaboration], Eur. Phys. J. C17, 207 (2000), DOI 10.1007/s100520000449.
\bibitem{PYTHIA} T.~Sjostrand,
Comput.\ Phys.\ Commun.\  {\bf 82}, 74 (1994).
\bibitem{mickey} M.~Chiu for the PHENIX Collaboration, these proceedings.
\bibitem{STAR_supp} C.~Adler {\it et al.} [STAR Collboration], nucl-ex/0206011. 

\end{thebibliography}
\end{document}